\documentclass{Interspeech}

\interspeechcameraready 

\title{DeepGESI: A Non-Intrusive Objective Evaluation Model for Predicting Speech Intelligibility in Hearing-Impaired Listeners}

\author[affiliation={1}]{Wenyu}{Luo}
\author[affiliation={1}]{Jinhui}{Chen}

\affiliation{}{Wakayama University}{Japan}
\email{luo.wenyu@g.wakayama-u.jp}

\usepackage{comment}
\usepackage{balance}

\begin{document}
\maketitle

\begin{abstract}
Speech intelligibility assessment is essential for many speech-related applications.

However, most objective intelligibility metrics are intrusive, as they require clean reference speech in addition to the degraded or processed signal for evaluation. Furthermore, existing metrics such as STOI are primarily designed for normal-hearing listeners, and their predictive accuracy for hearing-impaired speech intelligibility remains limited. On the other hand, the GESI (Gammachirp Envelope Similarity Index) can be used to estimate intelligibility for hearing-impaired listeners, but it is also intrusive, as it depends on reference signals. This requirement limits its applicability in real-world scenarios.
    
To overcome this limitation, this study proposes DeepGESI, a non-intrusive deep learning–based model capable of accurately and efficiently predicting the speech intelligibility of hearing-impaired listeners without requiring any clean reference speech.

Experimental results demonstrate that, under the test conditions of the 2nd Clarity Prediction Challenge(CPC2) dataset, the GESI scores predicted by DeepGESI exhibit a strong correlation with the actual GESI scores. In addition, the proposed model achieves a substantially faster prediction speed compared to conventional methods.
\end{abstract}

\section{Introduction}


Speech intelligibility assessment constitutes a fundamental aspect of modern speech technology. It serves not only to quantify how effectively speech enhancementalgorithms restore intelligibility under adverse acoustic conditions such as noise, reverberation, or distortion \cite{li2021multi}, but also to evaluate how hearing aids and cochlear implants improve speech understanding for individuals with hearing loss \cite{kates2022overview}. In addition, intelligibility metrics are crucial for analyzing how signal degradations affect recognition performance in automatic speech recognition (ASR) systems \cite{tu22b_interspeech}, and for examining whether converted speech in voice conversion \cite{wang23qa_interspeech} tasks preserves the linguistic clarity and comprehensibility perceived by human listeners.


A straightforward approach to estimating speech intelligibility is to conduct subjective listening tests in which speech samples are presented to human listeners, and intelligibility is quantified as the ratio of the number of correctly recognized words to the total number of words in the presented samples. However, conducting such tests to achieve reliable results demands extensive experimental resources and participant involvement, which considerably limits their feasibility in large-scale or real-world applications.

Accordingly, various objective methods have been proposed as alternatives to subjective listening tests for estimating speech intelligibility, such as the Speech Intelligibility Index (SII) \cite{american1997methods}, the Short-Time Objective Intelligibility (STOI) \cite{5713237}, and its extended version (ESTOI) \cite{7539284}. These metrics have been shown to exhibit strong correlations with subjective intelligibility scores under specific acoustic conditions.

Although these methods demonstrate high correlations with human intelligibility scores, they are intrusive, requiring clean reference speech in addition to the degraded or processed signal for evaluation. This dependence on reference signals limits their practicality in real-world scenarios, where clean speech data are not always available. On the other hand, non-intrusive methods estimate perceived speech intelligibility directly from the degraded or processed speech without relying on clean reference signals. An example of such an approach is the Non-Intrusive Short-Time Objective Intelligibility (NI-STOI) \cite{andersen2017non}. In recent years, deep learning based non-intrusive models have achieved remarkable progress in the assessment of speech intelligibility. These models are trained by minimizing the loss between the predicted and the ground-truth intelligibility scores, enabling them to estimate speech intelligibility without the need for clean reference speech.

These methods have substantially advanced speech intelligibility assessment, yet the majority of existing work still centers on normal-hearing (NH) listeners, while research specifically addressing intelligibility assessment for hearing-impaired (HI) listeners remains comparatively scarce.

The Gammachirp Envelope Similarity Index (GESI) \cite{irino22_interspeech}, which is built on the same framework as Gammachirp Envelope Distortion Index(GEDI) \cite{yamamoto2020gedi} and implemented using the gammachirp auditory filterbank (GCFB) \cite{irino2023hearing} together with modulation-frequency analysis via an MFB \cite{jorgensen2011predicting}, provides an objective method that can be applied to evaluating speech intelligibility for hearing-impaired listeners. A notable advantage of GESI is that it can account for hearing-loss effects by incorporating audiogram-dependent processing and modulation-domain cues, enabling more accurate prediction of intelligibility for HI listeners than conventional metrics. However, GESI still faces several limitations that hinder its use in practical settings. As an intrusive metric, it requires clean reference speech, which is often unavailable in real-world environments. In addition, its computational cost is relatively high due to the complex processing involved in the GCFB, modulation filterbank, and envelope-similarity stages, making real-time or embedded deployment difficult. In this study, we propose DeepGESI, a deep learning–based non-intrusive objective evaluation model designed to predict speech intelligibility for hearing-impaired listeners. DeepGESI achieves prediction accuracy comparable to that of GESI while operating in a fully non-intrusive manner, requiring no clean reference speech. In addition, its lightweight model design enables efficient computation, making it suitable for real-time applications.

The remainder of this paper is organized as follows. Section 2 describes the proposed DeepGESI model. Section 3 presents the experimental setup and results. Finally, Section 4 concludes this work.

\section{DeepGESI}

\subsection{Architecture}

\begin{figure}[t]
  \centering
  \includegraphics[width=0.8\linewidth]{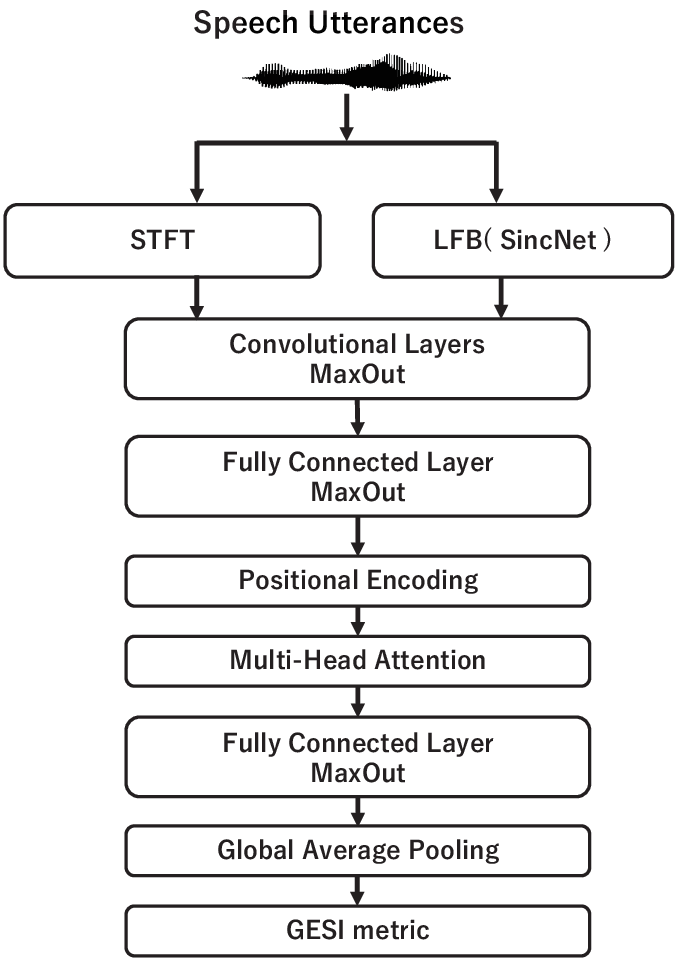}
  \caption{Architecture of the DeepGESI model.}
  \label{fig:Networkarchitecture}
\end{figure}

Figure \ref{fig:Networkarchitecture} shows the overall architecture of the proposed DeepGESI model. DeepGESI extracts two types of acoustic features using the short-time Fourier transform (STFT) and a learnable filterbank (LFB), after which the features are temporally aligned along the time axis. These two feature sequences are then jointly fed into the first convolutional layer, which integrates them into a unified time-series representation.

The output of the convolutional layer is passed to an attention module \cite{vaswani2017attention}, enabling the model to capture context-dependent salient information and long-range dependencies. After attention processing, frame-level GESI metric values are estimated through a fully connected layer. Finally, a global average pooling layer aggregates the frame-level predictions to produce the utterance-level GESI metric as the final output.

In summary, DeepGESI performs intelligibility estimation through a combination of local frame-level computations and global utterance-level aggregation, enabling effective prediction of the GESI metric.
\subsubsection{Auditory front-end feature extractor}
DeepGESI takes two types of acoustic features as input: spectral features extracted via the short-time Fourier transform (STFT) and learnable filterbank (LFB) features. In this study, the learnable filterbank is implemented using a SincNet-based design \cite{ravanelli2018speaker}, allowing the filter parameters to be optimized directly from data.

In contrast to conventional 1D convolution, which learns the kernel coefficients directly, SincNet keeps the kernel length fixed and learns two parameters, the lower cutoff frequency  $f_1$ and the upper cutoff frequency  $f_2$ , which define the passband. The filter shape is then determined from these parameters. This enables the model to more effectively capture narrow-band components of the speech signal, such as pitch and formant information.

In SincNet, the impulse response shown in equation \ref{eq:SCfilter} is obtained by taking the inverse Fourier transform of the ideal rectangular band-pass filter.
\begin{equation}
\label{eq:SCfilter}
g[n,f_1,f_2] = 2f_2\,\mathrm{sinc}(2\pi f_2 n) - 2f_1\,\mathrm{sinc}(2\pi f_1 n)
\end{equation}
Where $n$ denotes the time index over the kernel length $L$, and $f_1$ and $f_2$ are learnable parameters representing the lower and upper cutoff frequencies of the passband, respectively. Because the use of the sinc function results in a smoothly varying filter shape, gradient-based optimization becomes more stable \cite{ravanelli2018speaker}, leading to improved convergence.

\subsubsection{Maxout}

\begin{figure}[t]
  \centering
  \includegraphics[width=0.8\linewidth]{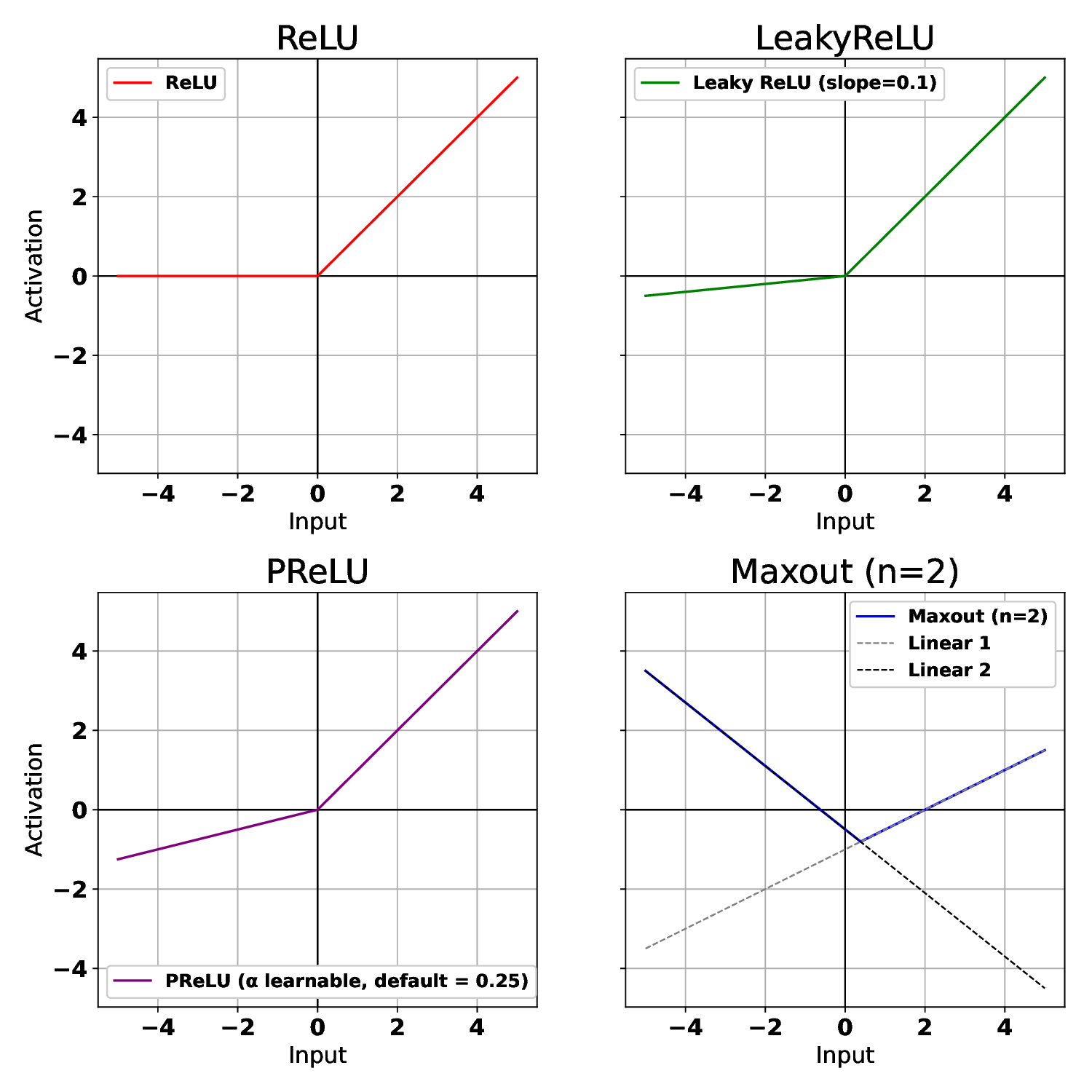}
  \caption{Mathematical characteristics of ReLU, LeakyReLU, PReLU, and Maxout.}
  \label{fig:maxout_relu}
\end{figure}

DeepGESI uses the Maxout activation function \cite{goodfellow2013maxout} instead of the commonly used ReLU \cite{nair2010rectified}. As shown in the upper-left panel of Figure~\ref{fig:maxout_relu}, ReLU outputs positive values as they are while mapping all negative values to zero. However, in speech analysis tasks, informative acoustic cues may also lie in the negative range, and applying ReLU discards such information, making it difficult for a lightweight model to learn these features and often requiring a more complex architecture to compensate. A common alternative is to use ReLU variants such as LeakyReLU \cite{maas2013rectifier} or PReLU \cite{he2015delving}, which preserve part of the negative range, but these approaches still have limitations.

On the other hand, as shown in the lower-right panel of Figure~\ref{fig:maxout_relu}, Maxout selects the maximum value among several linear transformations, allowing the activation function to adapt its shape dynamically according to the training data and thereby improving its ability to represent nonlinear features. The Maxout activation is defined as
\begin{equation}
    \mathrm{Maxout}(X) = \max_{i = 1, \dots, n} \left( \mathbf{w}_i^{\mathrm{T}} X + b_i \right)
\end{equation}
where $X$ denotes the input vector. Maxout applies multiple linear transformations, $\mathbf{w}_i^{\mathrm{T}} X + b_i$, to the input and outputs the maximum value among them as the activation.

\subsubsection{Positional Encoding}
Because the attention mechanism processes the entire input sequence as a set of vectors without any inherent notion of temporal order, it cannot preserve sequential or time-dependent relationships on its own. Therefore, to enable the model to exploit the positional structure of the sequence, additional information reflecting the relative or absolute positions of the input frames must be incorporated. There are many choices of positional encodings,learned and fixed\cite{vaswani2017attention, gehring2017convolutional}.In this work, we use Rotary Position Embedding (RoPE) \cite{su2024roformer} for positional encoding. Unlike learned positional embeddings \cite{gehring2017convolutional}, which require manually specifying the positional dimension according to the input length and may fail when an utterance of an unexpected length is provided , RoPE offers better extrapolation ability and can handle sequences of arbitrary length. This property allows the model to process input speech of varying durations without encountering length-related issues. RoPE applies a rotational positional encoding to the query and key vectors, defined as
\begin{equation}
  q_t^{\mathrm{rope}} = R(t)\, q_t
\end{equation}
\begin{equation}
  k_t^{\mathrm{rope}} = R(t)\, k_t
\end{equation}
where $R(t)$ is a block-diagonal rotation matrix composed of $d/2$ two-dimensional rotation blocks:
\begin{equation}
  R(t) =
  \mathrm{diag}\bigl(
    R(t\theta_0),\,
    R(t\theta_1),\,
    \ldots,\,
    R(t\theta_{d/2-1})
  \bigr)
\end{equation}
Where $d$ denotes the hidden dimensionality of the model, which is assumed to be even so that it can be decomposed into $d/2$ two-dimensional subspaces. And each $2\times2$ rotation block is defined as
%
\begin{equation}
  R(t\theta_i)=
  \begin{pmatrix}
    \cos(t\theta_i) & -\sin(t\theta_i) \\
    \sin(t\theta_i) &  \cos(t\theta_i)
  \end{pmatrix}
\end{equation}
\begin{equation}
  \theta_i = 10000^{-2i/d}
\end{equation}
RoPE rotates each two-dimensional subspace $(2i,\,2i{+}1)$ by an angle proportional to its position $t$, introducing 
absolute positional information into the representations.

Moreover, in the attention score 
$\langle q_t^{\mathrm{rope}},\, k_s^{\mathrm{rope}} \rangle$, 
where $t$ and $s$ denote the position indices of the query and key tokens, respectively,
the rotation satisfies $R(t)^{\top}R(s)=R(s-t)$,
so the attention score depends only on the relative displacement $(s-t)$. 
In addition, by following the frequency design of the sinusoidal encoding
$\theta_i = 10000^{-2i/d}$,
RoPE inherits its long-term decay property, meaning that the inner product naturally decreases
as the relative position increases.
This allows RoPE to encode absolute positions in the representations while incorporating
relative positional information in the attention computation.

\subsection{Objective function}

We treat speech intelligibility prediction as a regression task and construct the loss function using the least-squares criterion. Under the assumption of Gaussian noise, least squares is equivalent to maximizing the likelihood and is therefore widely used in regression analysis. In an utterance, both temporally stationary and non-stationary noise components may be present at each frame. Consequently, when predicting intelligibility, it is necessary to account not only for global acoustic characteristics over the entire utterance but also for local, frame-level acoustic features. Accordingly, DeepGESI adopts the following loss function:
\begin{equation}
    L = L_{\mathrm{sent}} + \alpha L_{\mathrm{frame}}
\end{equation}
Where$L_{\mathrm{sent}}$ denotes the loss computed from the global prediction error at the utterance level, whereas $L_{\mathrm{frame}}$ represents the loss derived from the local prediction error at each frame. The weighting coefficient $\alpha$ is a hyperparameter that controls the balance between the two terms. The definitions of $L_{\mathrm{sent}}$ and $L_{\mathrm{frame}}$ are given as follows:
\begin{equation}
    L_{\mathrm{sent}}=\frac{1}{B}\sum_{i=1}^{B}\left(y_i-\widehat{y_i}\right)^2
\end{equation}
\begin{equation}
    L_{\mathrm{frame}}=\frac{1}{B}\sum_{i=1}^{B}\left(\frac{1}{T_i}\sum_{t=1}^{T_i}\left(y_i-{\hat{f}}_{i,t}\right)^2\right)
\end{equation}
where $\widehat{y}_{i}$ is the predicted sentence-level DeepGESI metric for the $i$-th training sample, and $y_{i}$ is the corresponding ground-truth GESI metric. The value $T_{i}$ denotes the number of frames in the $i$-th sample, and $\hat{f}_{i,t}$ represents the predicted frame-level DeepGESI metric at frame $t$. The variable $B$ is the batch size used during training. In this study, the weighting coefficient was set to $\alpha = 1$.

\section{EXPERIMENTS}

\subsection{Dataset}

In this study, we use the hearing-aid output signals from the publicly available CPC2 (The 2nd Clarity Prediction Challenge) dataset \cite{barker20242nd} as the material for model training and evaluation. The dataset is generated from complex indoor scenes based on CEC2 \cite{akeroyd20232nd}and contains speech signals processed by various hearing-aid algorithms under diverse speech-in-noise conditions, covering different noise types, speakers, and interfering source configurations.

\subsection{Experimental Setup}

To train DeepGESI, we used all 5,946 hearing-aid output signals provided in the CPC2 training dataset. These signals were randomly divided into training, validation, and test sets using an 80\%, 10\%, and 10\% split, respectively. The model was trained to minimize the prediction error with respect to the GESIv123 \cite{yamamoto2023gesi} metric, which in this study was computed using parameters corresponding to an 80-year-old simulated hearing loss within the frequency range of 125~Hz to 8~kHz. All audio signals were downsampled to 16~kHz to improve training efficiency.

In this study, the verification experiments on accuracy and computational speed were conducted on a PC equipped with an AMD Ryzen~5~5600X (6~cores) and an NVIDIA GeForce RTX~4070 GPU. For training DeepGESI, the Adam optimizer was used with a learning rate of \(1\times10^{-4}\) and a batch size of~6.


Three metrics were used for evaluation: MSE, LCC, and SRCC. Lower MSE indicates better prediction accuracy, while higher LCC and SRCC indicate stronger agreement with the ground truth.

\subsection{Detailed assessment results}

\subsubsection{Evaluation on Seen and Unseen Datasets}

In this study, the 10\% test split from the above partition was used as the seen evaluation set. In addition, the CPC2 dataset provides an official evaluation set consisting of 897 signals, which we treat as the unseen evaluation set. The unseen data include speakers, interference conditions, and acoustic characteristics that do not appear in the training or validation process, making them acoustically distinct from the seen data.
\begin{figure}[t]
  \centering
  \includegraphics[width=0.49\linewidth]{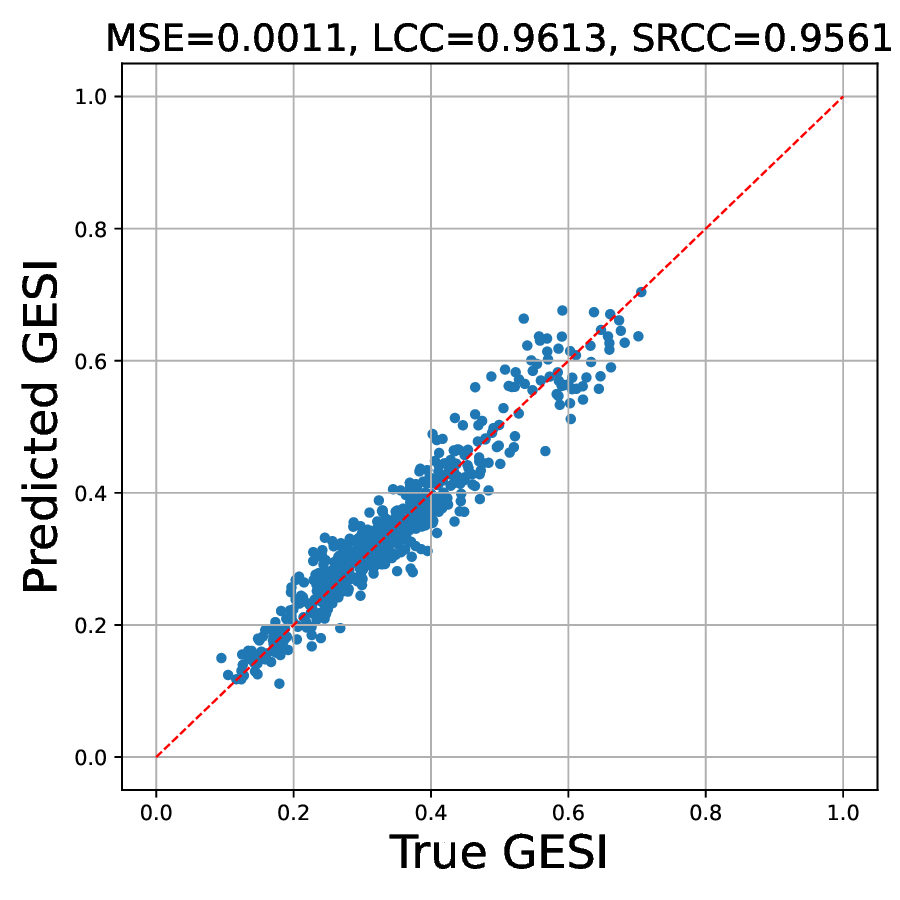}%
  \hfill
  \includegraphics[width=0.49\linewidth]{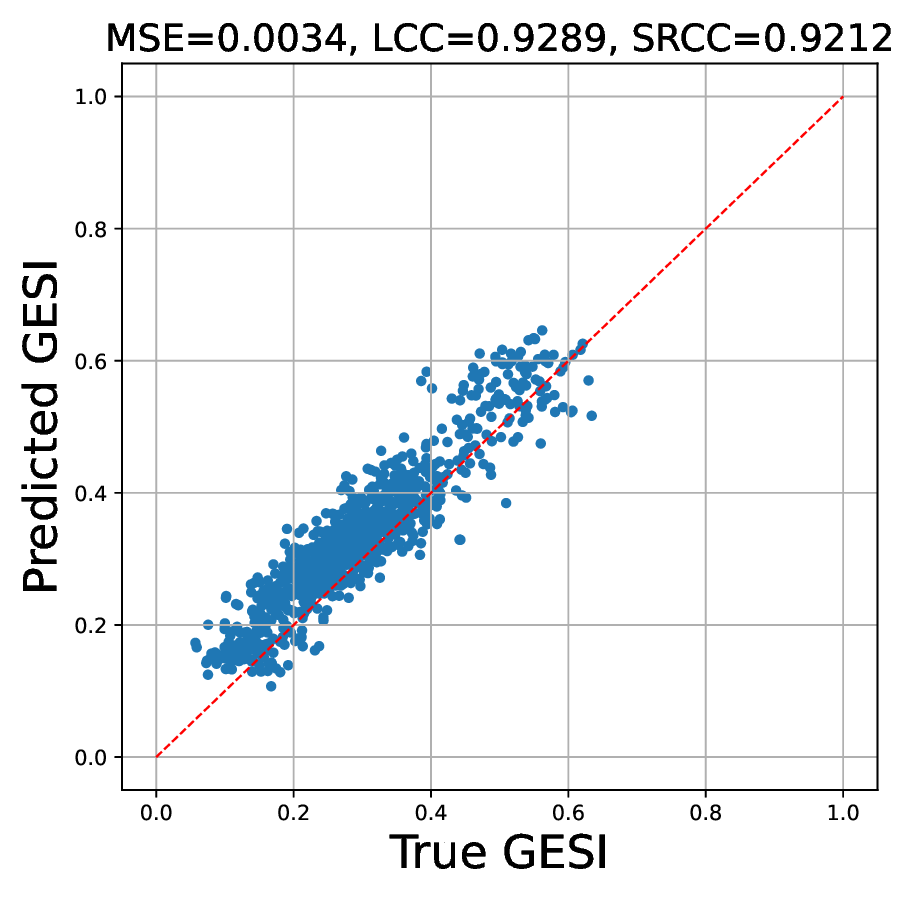}
  \caption{Scatter plots of speech intelligibility assessment by DeepGESI under the seen (left) and unseen (right) conditions.}
  \label{fig:twofigs}
\end{figure}
%



Figure~\ref{fig:twofigs} (left) shows that DeepGESI achieves MSE = 0.0011, LCC = 0.9613, and SRCC = 0.9561 on the seen evaluation set, indicating strong agreement with the ground-truth GESI metrics on the CPC2 dataset.

Similarly, as shown in Figure~\ref{fig:twofigs} (right), the unseen evaluation set yields MSE = 0.0034, LCC = 0.9289, and SRCC = 0.9212. Although the correlations are slightly lower than those on the seen data, the performance remains stable, demonstrating strong generalization under unseen acoustic conditions.

\subsubsection{Effect of Activation Functions}
\begin{table}[t]
  \caption{Performance comparison of activation functions on seen and unseen data.}
  \label{tab:activation_functions}
  \centering
  \resizebox{\linewidth}{!}{
  \begin{tabular}{l l c c c}
    \toprule
    \textbf{Data} & \textbf{Activation Functions} & \textbf{MSE} & \textbf{LCC} & \textbf{SRCC} \\
    \midrule
    Seen   & Maxout     & \textbf{0.0011} & \textbf{0.9613} & \textbf{0.9561} \\
           & ReLU       & 0.0016 & 0.9429 & 0.9305 \\
           & LeakyReLU  & 0.0016 & 0.9435 & 0.9337 \\
           & PReLU      & 0.0022 & 0.9238 & 0.9073 \\
    \midrule
    Unseen & Maxout     & \textbf{0.0034} & \textbf{0.9289} & \textbf{0.9212} \\
           & ReLU       & 0.0040 & 0.9092 & 0.8913 \\
           & LeakyReLU  & 0.0041 & 0.9135 & 0.8987 \\
           & PReLU      & 0.0042 & 0.8924 & 0.8800 \\
    \bottomrule
  \end{tabular}
  }
\end{table}
%
%
Table~\ref{tab:activation_functions} presents the performance comparison of activation functions on the seen and unseen evaluation sets. Compared with ReLU and its variants, the Maxout activation function shows superior performance in this task. This confirms that Maxout is more effective at representing nonlinear acoustic features, which is particularly beneficial in speech analysis tasks.

\subsubsection{Effect of Positional Encoding}
\begin{table}[t]
  \caption{Performance of RoPE (Rotary Position Embedding), Sinusoidal PE (Sinusoidal Positional Encoding), and LPE (Learned Positional Embedding) on seen and unseen data.}
  \label{tab:positional_encoding}
  \centering
  \resizebox{\linewidth}{!}{
  \begin{tabular}{l l c c c}
    \toprule
    \textbf{Data} & \textbf{Positional Encoding} & \textbf{MSE} & \textbf{LCC} & \textbf{SRCC} \\
    \midrule
    Seen   & RoPE              & \textbf{0.0011} & \textbf{0.9613} & \textbf{0.9561} \\
           & Sinusoidal PE     & 0.0015 & 0.9464 & 0.9344 \\
           & LPE               & 0.0037 & 0.8632 & 0.8430 \\
    \midrule
    Unseen & RoPE              & \textbf{0.0034} & \textbf{0.9289} & \textbf{0.9212} \\
           & Sinusoidal PE     & 0.0038 & 0.9213 & 0.9087 \\
           & LPE               & 0.0059 & 0.8039 & 0.7802 \\
    \bottomrule
  \end{tabular}
  }
\end{table}
%

Table~\ref{tab:positional_encoding} shows the performance of RoPE, sinusoidal positional encoding, and learned positional embedding on the seen and unseen data. Compared with sinusoidal positional encoding, RoPE achieves better generalization performance. This improvement may be related to the long-term decay property induced by the frequency design, which aligns with the intuition that distant tokens should have weaker interactions. In comparison with learned positional embedding, RoPE can naturally handle input sequences of arbitrary length, whereas learned positional embedding requires a manually specified maximum length in advance, and the choice of this predefined length can affect performance.

\subsubsection{Evaluation of Computational Speed}
DeepGESI was compared with GESI \cite{yamamoto2023gesi} and HASPI \cite{kates2021hearing} in terms of computational speed. To evaluate processing time, 500 speech signals were randomly selected from the CPC2 unseen dataset, which contains 897 signals in total, and all experiments were conducted under the same PC environment. 

\begin{figure}[t]
  \centering
  \includegraphics[width=1.0\linewidth]{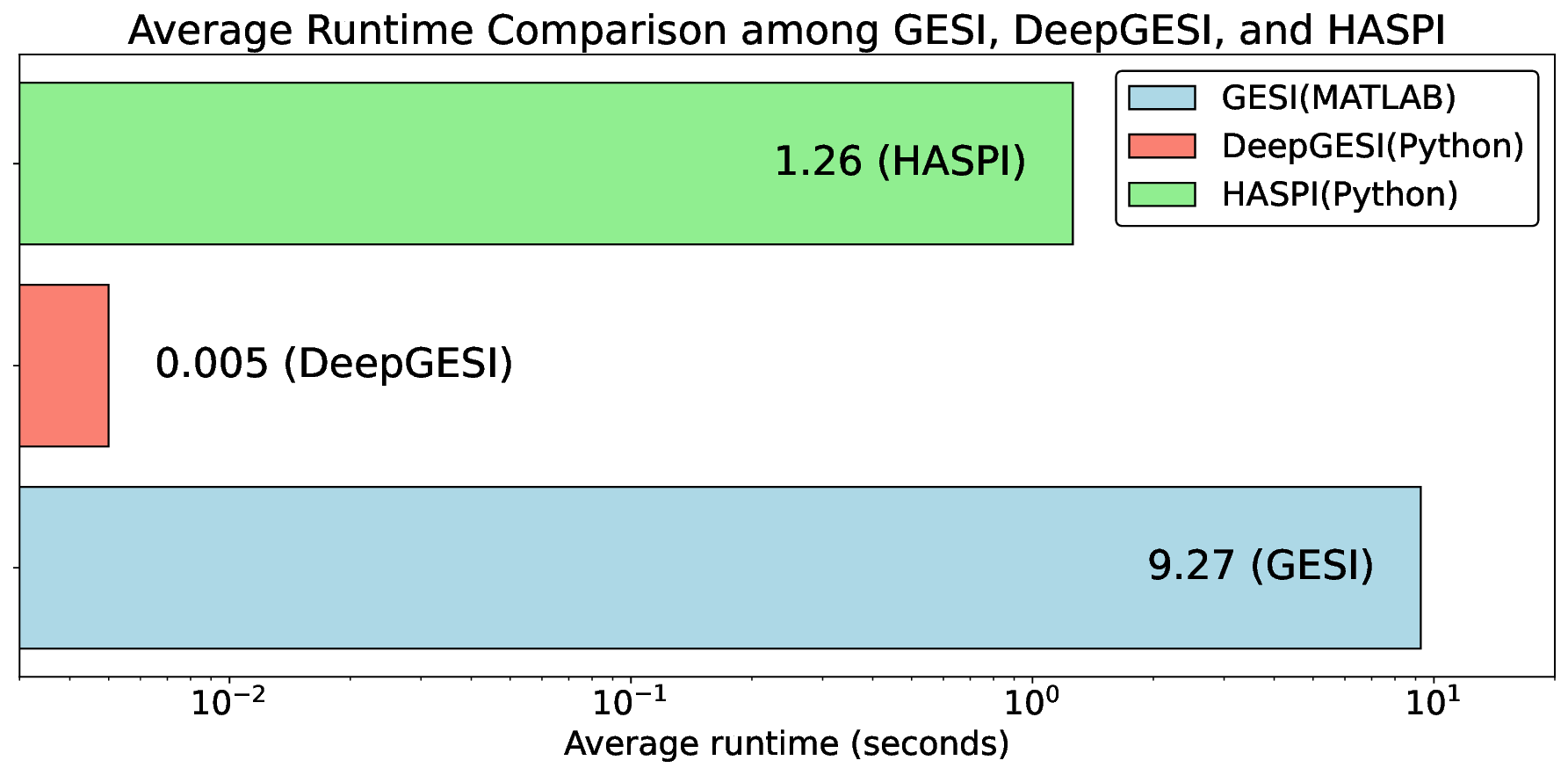}
  \caption{Computation time per utterance for each method.}
  \label{fig:runtime}
\end{figure}

As shown in Figure~\ref{fig:runtime}, DeepGESI requires on average 0.005~seconds per utterance, whereas GESI and HASPI require 9.27~seconds and 1.26~seconds, respectively. Compared with these conventional methods, DeepGESI is significantly faster, confirming that it can estimate speech intelligibility for evaluation signals in real time.
\section{Conclusion}

In this study, we proposed DeepGESI, a deep learning--based model for estimating speech intelligibility for hearing-impaired listeners that overcomes the limitations of the conventional intrusive method GESI. DeepGESI achieves performance comparable to GESI while operating in a non-intrusive manner and enabling real-time processing. 

In the present work, DeepGESI was not fine-tuned using subjective listening test results, which remains an important direction for future work. Further evaluations under more diverse acoustic conditions will be conducted, and the model and loss function will be refined based on additional experimental.




\clearpage
\enlargethispage{-19\baselineskip}
\bibliographystyle{IEEEtran}
\bibliography{mybib}

\end{document}